\documentclass[10pt,conference]{IEEEtran}
\usepackage[utf8]{inputenc}
\usepackage[T1]{fontenc}
\usepackage{subcaption}
\usepackage{hyperref}
\usepackage{graphicx}
\usepackage{multirow}

\def\code#1{\texttt{#1}}

\title{\huge The standard coder: a machine learning approach to
measuring the effort required to produce source code change}

\author{\IEEEauthorblockN{{Ian Wright}
\IEEEauthorblockA{Semmle Inc.\\
Blue Boar Court\\9 Alfred Street\\Oxford OX1 4EH, UK\\
Email: wright@semmle.com}}
\and
\IEEEauthorblockN{Albert Ziegler}
\IEEEauthorblockA{Semmle Inc.\\
Blue Boar Court\\9 Alfred Street\\Oxford OX1 4EH, UK\\
Email: albert@semmle.com}}

\begin{document}

\flushbottom
\maketitle
%
%
\thispagestyle{empty}

\begin{abstract}
We apply machine learning to version control data
to measure the quantity of effort required to produce source code changes. We construct a  
model of a `standard coder' trained from examples of code changes
produced by actual software developers together with the labor time they
supplied. The effort of a code change is then defined as the 
labor hours supplied by the standard coder to produce that change. We therefore reduce heterogeneous, structured code changes to a scalar measure of effort derived from large quantities of empirical data on the coding behavior of software developers. The standard coder 
replaces traditional metrics, such as lines-of-code or function point
analysis, and yields new insights into what code changes require more
or less effort.
\end{abstract}

\section{Introduction}

Software developers create software by incrementally changing
a codebase. Some changes are easy, and therefore
require little effort; other changes are more difficult,
and require more effort. For example, developers know, from experience,
that globally renaming a variable is almost effortless, whereas refactoring a class hierarchy is harder. Can we quantify such intuitions? 
Are there laws or regularities that relate
{\em kinds} of changes to the {\em amount} of effort
required to produce them?

The advent of large-scale, open-source development that
manages code change with version control systems has
generated large volumes of publicly available time-series
of aspects of the software development process. In particular,
version control systems associate source code changes with commit 
timestamps. Hence, we have examples
of code changes associated with highly noisy time intervals during which 
the changes were typically produced. 

This data implicitly contains insights into the relationship
between kinds of code change and their effort. For instance,
if we can (a) capture the salient properties of code change
that affect effort, and (b) infer the actual 
coding time supplied by developers to make those changes, then (c) we can
train a model, using machine learning techniques, that 
predicts the typical time required by a generic developer to make any code change,
including code changes not represented in the training data.
Such a model would capture the principal statistical regularities 
between code change and effort. In effect, the model 
represents a `standard coder' that encapsulates the typical
coding behavior of a large population of software developers.
We would hope, for example, that the standard coder 
takes less time to rename a variable (e.g., 10 minutes), and 
more time to refactor a class hierarchy (e.g., 3 hours), and so on.

Existing methods of measuring effort rely on overly simple metrics, such as lines-of-code change (\cite{LOCasEffort}, \cite{LOCasEffort2}), subjective methodologies (e.g., function point analysis \cite{functionPointAnalysis79}) or models constructed from small datasets (e.g., COCOMO
\cite{cocomo2000}) that require manual data collection. Instead we aim to automatically and non-invasively measure effort from a side-effect of software engineering (i.e., version control data). Also, measuring effort in terms of a standard coder is, in principle, maximally unbiased, especially compared to metrics with {\em a priori} assumptions,
since the standard coder is empirically based on the actual behavior of a 
large community of software developers.

We need to measure software engineering effort to gain
a deeper understanding of the practice of software development
and to improve it. For example, a standard coder
could tell us what kinds of code changes are
typically effortless or effortful in order to identify
production bottlenecks. We could compare
developer output (measured in coding hours supplied by the standard coder) with the labor time actually supplied in
order to understand the relationships between 
developer productivity and factors such
as (a) problem domain (e.g., user interfaces versus
business logic), (b) developer experience (e.g., beginner versus
expert) and (c) the quality of a codebase 
(e.g., whether easy or hard to maintain), and so on.

The structure of this paper is as follows: (i) we
describe our dataset, (ii) we explain
how we control for noise in commit intervals, (iii)
we define the standard coder, (iv) we discuss 
some preliminary insights, and (v) we conclude.

\section{The dataset}

Our data was collected from Semmle's public code analysis tool, LGTM.com
(`looks good to me'). We fetched source code changes and commit timestamps for $\approx 500$K commits submitted by $\approx 20$K developers to $\approx 5$K open source projects. Each commit produces a new source code `snapshot' that LGTM analyzes using QL \cite{QL2016}, a declarative, object-oriented logic programming language. QL treats `code as data' and is particularly well suited for static code analysis. LGTM filters out source code files that do not represent work performed, such as inclusion of third-party library code, generated or minified code etc. For each commit we fetch the author time-stamp of the commit (i.e., the time when the code change was committed on the developer's local machine) and the modified files and their corresponding prior versions in the parent commit (if they exist). Hence, our raw data consists of timestamped commits together with before and after versions of source code files altered by the commit.

\section{Mining coding time from commit intervals}

Consider a parent and child commit separated by a clock-time interval of $x$ hours. The actual coding work occurs sometime in this interval (with rare exceptions). But this interval is a very noisy estimate of coding time since developers are not always coding, but spend time on other tasks either indirectly related to coding, or not related at all. In order to associate code changes with effort we must reduce this noise and infer a best-guess estimate of the actual coding time supplied.

\begin{figure}
  \centering
  \includegraphics[width=0.9\linewidth]{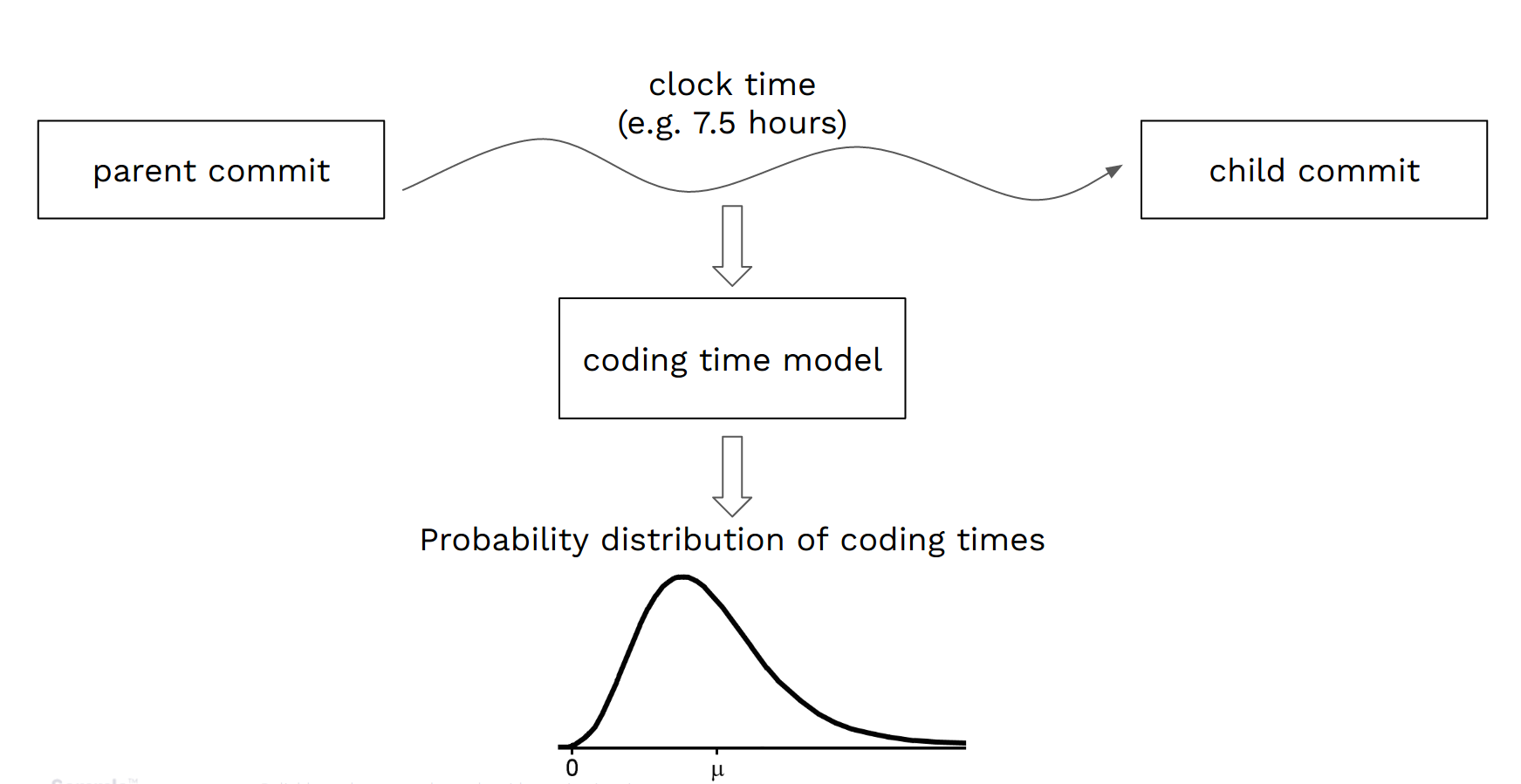}
  \caption{\small {\em The coding time model}: commits are timestamped so we know the clock time duration between a parent and child commit. But developers aren't always coding. So this interval is a very noisy estimate of coding time. We convert clock time to a probability distribution of coding times using a neural hidden Markov model trained on the specific commit patterns of individual developers.}
  \label{fig:mining-coding-times}
\end{figure}

Is this possible? Consider examining the commit time-series data for
a specific developer. We might notice certain regularities: for example, they never work weekends. So, with some probability, we can subtract weekends from their clock times in order to get closer to the true coding time supplied. But the data will contain additional hints: for example, we might notice they tend to commit between certain hours of the day, such as between 9am and 6pm. So again, with some probability, we can ignore the time spent outside their typical working hours. 
Also, given our knowledge of how developers actually work, then a large number of contiguous commits close in time (e.g, minutes) probably indicates a sustained period of uninterrupted coding, whereas commits far apart in time probably indicate independent bursts of activity. 
In summary, an individual developer's commit pattern is a noisy manifestation of their underlying coding time behavior.

Our strategy, then, is to construct a probabilistic {\em coding time
model} for each developer that captures the most likely `hidden' coding behavior that generates the observed time-series of their commit events. We then use the model to infer the probability of their coding activity during commit intervals (see Figure \ref{fig:mining-coding-times}). In essence, we map (noisy) commit intervals to (less noisy) coding times by learning a probabilistic generative model with inductive bias that encodes heuristic facts about how developers produce code.


\subsection{A generative model of coding activity}

\begin{figure}
  \centering
  \includegraphics[width=0.9\linewidth]{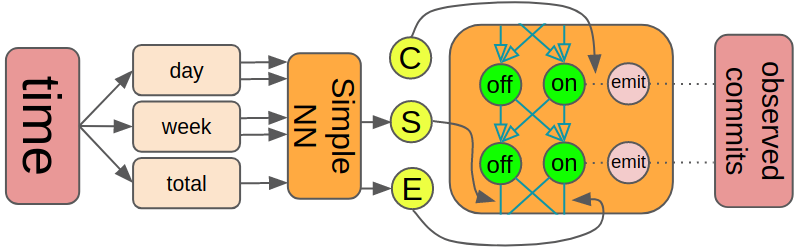}
  \caption{\small The NN computes time dependent transition probabilities $S$ and $E$ for the HMM. The HMM's state probabilities  are fitted to the observed sequence of commits with the forward-backward algorithm. The NN and emission probability $C$ are trained to maximize the likelihood of the observed sequence.}
  \label{fig:neuralHMM}
\end{figure}

We assume a developer is in one of two states: either coding or not-coding. 
We define `coding' as an activity that can generate a commit event (e.g., coding at a keyboard and monitor) and `not-coding' as an activity that cannot (e.g., not in a workplace, or in the workplace but performing a non-coding activity). We observe the timestamped sequence of commit events: a commit can only happen when coding (with a trained probability $C$ per minute).

The transition from coding to not-coding occurs with a probability $E(t)$ (for end coding), and from not-coding to coding with probability $S(t)$ (for start coding). Hence we have the time-inhomogeneous state transition matrix:

\begin{table}[h!]
\centering
\begin{tabular}{lll}
                                & coding                      & not coding                  \\ \cline{2-3} 
\multicolumn{1}{l|}{coding}     & \multicolumn{1}{l|}{$1-E(t)$} & \multicolumn{1}{l|}{$E(t)$}   \\ \cline{2-3} 
\multicolumn{1}{l|}{not coding} & \multicolumn{1}{l|}{$S(t)$}   & \multicolumn{1}{l|}{$1-S(t)$} \\ \cline{2-3} 
\end{tabular}
\end{table}

In contrast to classical hidden Markov models (HMMs), probabilities $E$ and $S$ vary over time to account for habits (e.g., preponderance for evening coding) and schedule (e.g., only coding on week days). A simple neural network (NN) 
with 5 inputs --
the sine and cosine of the angle of an imaginary dial on a day-long and week-long clock, and normed overall time -- supplies the probability values. These features can encode daily or weekly patterns that may shift over time. The NN is then trained by back-propagating likelihood gradients through the HMM part (see Figure \ref{fig:neuralHMM}). We call the resulting architecture a neural hidden Markov model\footnote{Since the 90s, efforts have been made to combine the flexible power of neural networks with the sequential approach of HMMs. \cite{bengio91} used NN to generate the HMM’s outputs, while recently \cite{tran16} obtained excellent results by obtaining HMM parameters from NN outputs for parts-of-speech tagging.}.


We train a neural hidden Markov model to predict the probability 
an individual developer is currently coding for each one-minute interval of their recent history (e.g., $\approx$ 2 years). We then use the model to estimate the expected actual coding time between two successive commits.

\begin{figure}
  \centering
  \begin{subfigure}{.5\textwidth}
  \centering
  \includegraphics[width=\linewidth]{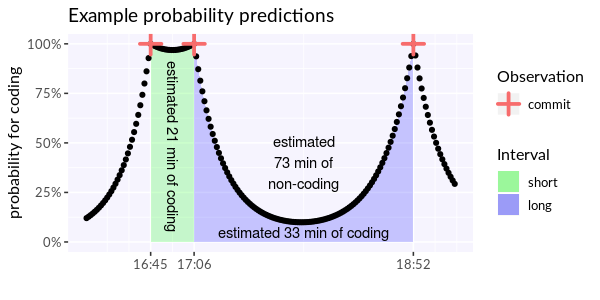}
  \caption{{\em The probability of coding during commit intervals.} HMM assigns a coding probability to each minute. Short commit intervals often indicate uninterrupted coding, whereas longer intervals suggest coding interruptions. We estimate time spent coding as the expectation value, which can be computed by taking the area under the probability curve.}
  \label{fig:probabilityPredictions}
  \end{subfigure}%
  \\
  \begin{subfigure}{.5\textwidth}
  \centering
  \includegraphics[width=\linewidth]{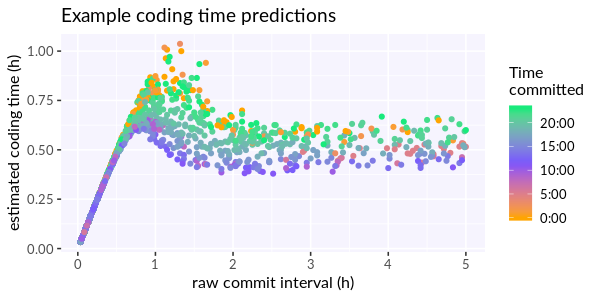}
  \caption{{\em Mapping commit intervals to estimated coding time.} Short intervals (e.g., $<1$ hour) are often spent coding nonstop. So, counter-intuitively, they often provide evidence for more coding time than medium intervals (e.g., 2 - 5 hours), which
  probably include coding breaks. The model has learned
that this developer has fewer interruptions during the morning than during the evening.}
  \label{fig:codingTimePredictions}
  \end{subfigure}
  \caption{Estimating coding time from commit intervals with a neural HMM.}
  \label{fig:predictions}
\end{figure}

The neural HMM is a flexible solution to a general problem: for any process that alternates between active and inactive phases, it infers the probability of being active at any time from an 
observed rhythm of potentially infrequent `life signs' (in this case commit events). The neural HMM can discover night-time coders, or developers that
code only on weekdays, or weekends (compare \cite{claes2018programmers}). 
The upshot is a model that maps commit intervals to estimates of coding time (see Figure
\ref{fig:predictions}).

\subsection{Validation of the coding-time model}

\subsubsection*{Simulated ground truth}

The main challenge when testing the coding time model is the lack of an observable ground truth. Gathering accurate information about which minutes of a developer’s life are spent coding would be invasive, time-consuming and error-prone. 

However, we can simulate the coding
behaviors of different developers and record both observable commit data and the normally hidden coding time data, and then validate the model against these simulations. This tells us (i) how much data and training time the model requires, (ii) how accurately the
model predicts individual commits, (iii) how robust the model is to controlled violations of its assumptions (e.g. if the distribution of the length of active and inactive time periods differs from our Markov assumptions)
and (iv) whether the the model detects certain patterns, e.g. a change in working schedule.

Under reasonable assumptions\footnote{A constant schedule of working 9--5 on weekdays with an exponential
distribution of the length of coding blocks and non-coding blocks with expectation values of 50 minutes in the morning and 20 minutes in the afternoon for coding, and 30 minutes for non-coding, and commit probability $4\%$ per minute while coding.} and a sample size of at least 100 commits, the correlation coefficient between actual coding time and neural HMM prediction exceeds $.69$ after at most $400$ epochs in all 20 experiments and never exceeded $0.75$. With a sample size of 50 commits, the correlation coefficient always exceeds $0.65$, but needs up to 800 epochs (data collection was every 100 epochs).

\begin{figure}
  \centering
  \includegraphics[width=0.9\linewidth]{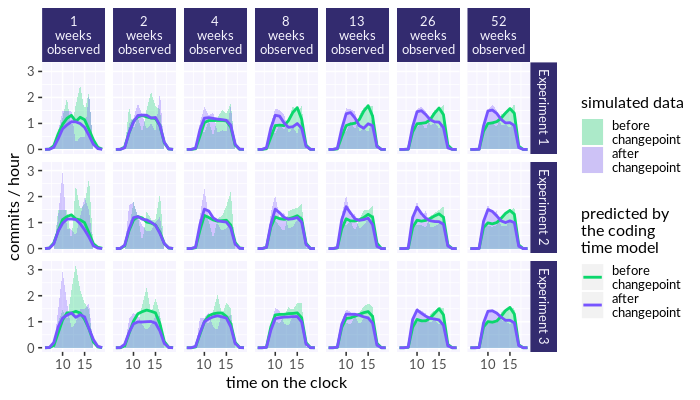}
  \caption{\small {\em Learning to recognize regime changes}: Simulated (shaded areas) and predicted (coloured  lines) commit probabilities at different times of the day before (green) and after (blue) a change in coding behavior. The model always learns the correct weekdays and office hours, but needs to observe roughly $8$--$13$ weeks of each coding pattern regime to learn the subtle difference between them.}
  \label{fig:regime_change}
\end{figure}

To measure the model's ability to detect change points, we simulate developers with a schedule change halfway through their recent history (Figure \ref{fig:regime_change}). For a number of $n$ weeks (green), they adhere to the schedule described above. Then for the same number of $n$ weeks, they adhere to the mirrored schedule with afternoon and morning coding specifications swapped. If $n$ is small (left side of the grid), it turns out that the model only learns the "average" schedule and predicts it to hold over the whole period of time (green and blue lines). With sufficient data ($n$ around $8$--$13$), however, the model learns two different schedules (green and blue lines) which are close to the actual exhibited ones (green and blue shaded areas).
To adequately learn this change point the model needed to observe roughly $8$ -- $13$ weeks of each regime (see Figure \ref{fig:regime_change}).

\subsubsection*{Prior versus posterior expectation}

Once the neural part of the neural HMM is trained to a developer's commit patterns, we can predict the chance that a developer
is coding by running the HMM in two different ways: (i)  \textit{live} mode that predicts only on past information, i.e. how long ago the last commit was, and (ii) \textit{hindsight} mode that predicts on both the past and the future, i.e. how long ago the last commit was \textit{and} how far into the future the next commit will be.

On switching from live to hindsight mode, the estimated chance of coding normally increases if the next commit is closer than expected, and decreases if further into the future than expected. The average change is a measure of the model's applicability for a developer. We will call the average change for a given chance of coding according to the live mode as the \textit{probability correction}.

For example, assume a developer tends to commit in more regular intervals than predicted by the Markov assumptions. Then the chance to be coding, according to live mode, is high typically soon after the last commit. However, since the developer's commits are more regular, the next commit arrives later than expected by the model. Upon switching to hindsight mode, and learning exactly when the next commit arrives, the model will correct the probability downwards. Conversely, the probability correction is positive for high coding probabilities.
We can therefore test the model's validity by checking whether the probability correction has expected value close to 0, independent of the model's current prediction.
%

For each author in our corpus with >$1000$ commits (for sufficient test data) we split their overall time period into 10 equal parts, resulting in between $32\times10^3$ and $11\times10^4$ data points. For each part and for each decile of posterior probability (according to live mode) we compare whether the average change is larger or smaller than 0, then perform a binomial significance test for a systematic tendency to under- or over-correct. We consider a test statistic of 0, 1, 9 or 10 to be a positive result, resulting in an expected false positive rate of $5.52\%$. The actual positive rate is $8\%$ ($24$ out of $300$), which is not statistically significantly different from the theoretical false positive rate (again using a binomial test).

We conclude that the model adequately captures the actual probabilities. 
In order to use all the available information we obtain these predictions in hindsight mode.

\begin{figure}
  \centering
  \includegraphics[width=0.9\linewidth]{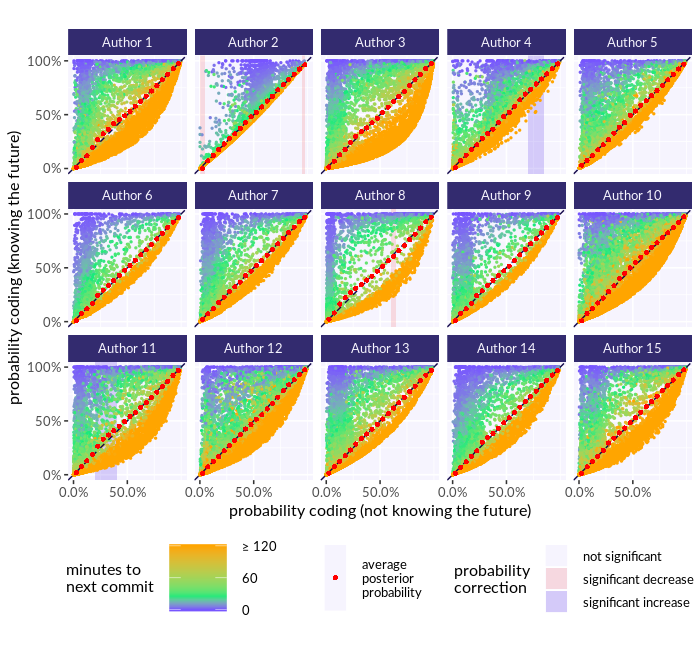}
  \caption{\small {\em Probability correction upon learning the future}: Each orange-to-blue point represents a one minute time interval for an author and the predicted probability for coding based on either knowing only the author's commits before that interval (x-axis) or knowing all commits (y-axis). If the time to next commit (blue for short, orange for long) is longer (resp. shorter) than expected by the model, the posterior probability will be lower (respectively higher) than the prior probability. The average posterior probability (red) is very close to the prior one. Differences with $p < 0.053$ have been indicated by a shaded background.}
  \label{fig:prior_posterior_prob}
\end{figure}

\section{The standard coder}

We use the coding time models, described in the previous section, to create datasets that associate code changes with estimated coding times. Next, we train a model that, given any code change, predicts the coding time typically required to produce it. The model then 
represents the coding time supplied 
to produce different kinds of code change by the `average' or representative {\em standard coder}.

How do we represent code changes, and what kind of regression model should we apply? On our datasets, a variant of bag-of-words features plus a deep mixture density network (MDN) yields the best performance.

A single commit may affect multiple files. For each language we construct a `token dictionary' consisting of separators (e.g., \verb|.|, \verb|(|, \verb|)|, \verb|+|, \verb|*|, etc.), keywords (e.g., \verb|if|, \verb|return|, \verb|while|, \verb|class|, etc.) and top $n$ most frequent words (e.g., popular variable names). For each touched file we compute the diff (lines either deleted and inserted) as per standard utilities (e.g., Unix's \verb|diff| command). We assume code deletion also requires coding effort. So we concatenate all diffs, both insertions and deletions, to form a composite change string. We compute a bag-of-words feature vector, which represents the quantity of token turnover, or code change, introduced by the commit. 

The relationship between code change and coding time is irreducibly noisy due to fundamental properties of the generative process (e.g., due to skill
differences between developers and random noise due to work
interruptions). In consequence, the regression model must be robust to noise. Plus, we
want to bound the uncertainty of our coding time predictions. A MDN \cite{MDNs} fulfills both criteria.
Deep MDNs are multi-layered neural networks that predict the parameters of a mixture distribution (i.e., predict a random, rather than deterministic, target), where the parameters are a nonlinear function of input features. We train with a likelihood loss. 

We trained a MDN model on 103K examples of Java
code changes (with commit intervals replaced by 200 samples each from
the estimated coding time distributions) resulting in $\approx$20M
training samples. The MDN predicts a mixture of 20 Gaussians using 116
bag-of-token features and 5 hidden layers of size 256,
64, 64, 64, and 64.\footnote{Note that the holdout
data is not drawn from the same distribution
as the training set. In consequence, at prediction time,
we truncate the mixture distribution so the
entire probability mass is contained within the interval
$[0,1]$ hour. The final prediction is the mean of
this truncated distribution.}

\begin{figure}[t!]
  \centering
  \includegraphics[width=0.84\linewidth]{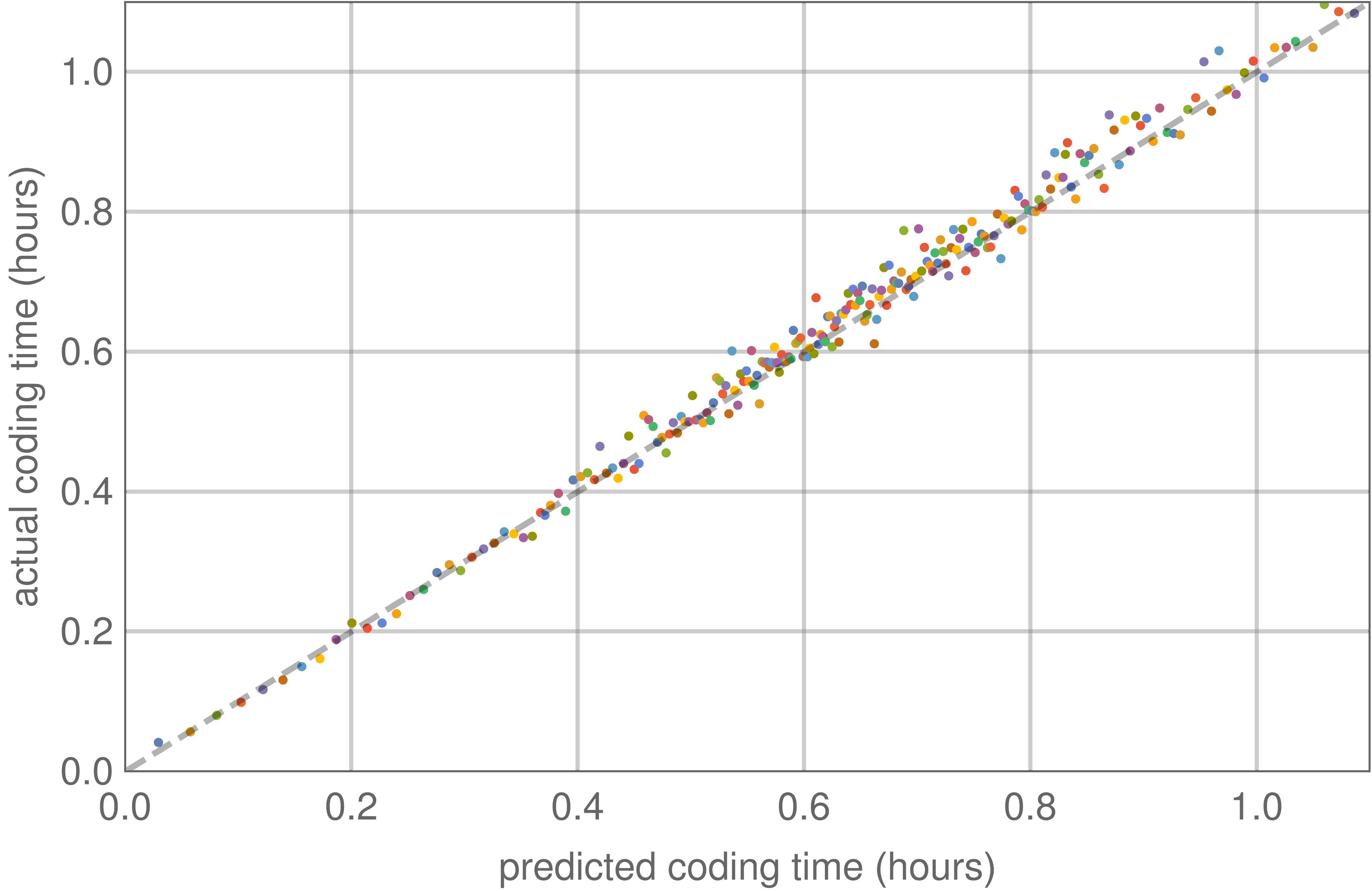}
  \caption{\small {\em Predicting the typical effort of code changes}. Each data point compares the predicted mean coding-time (x-axis, computed from timestamps) with the ground truth mean coding time (y-axis, computed from source code diffs) for binned hold-out data (250 bins each containing $\approx 4800$ Java code change examples). A high correlation, $R^2 = 0.99$, implies
  that the mixture density network captures the `average' or standard coding times of different kinds of code changes.}
  \label{fig:yyValidation}
 \end{figure}
 
The MDN learns distributions of coding times that correspond to similar kinds of code changes. To validate the model we need a ground truth. We know
that very short commit intervals
(e.g. $<1$ hour) typically correspond to
sustained periods of actual coding.\footnote{The coding time model
confirms this expectation since, for all developers,
it discovers a near one-one relation between coding times and short
commit intervals (see Figure \ref{fig:codingTimePredictions}).}
We therefore select, as our ground truth, holdout data with commit intervals $<1$ hour (and, to avoid any possible interference from the coding time model, we simply set coding times equal to the raw commit intervals).

We neither expect, nor require, 
that our model predict individual examples with high accuracy. Rather, we aim to predict the `average' coding
time of different groups of commits (selected by the model itself). So,
for model validation, we collect 
holdout examples into bins of similar predicted coding
times, and then compare against the bin means of actual coding times.
Figure \ref{fig:yyValidation} illustrates that
our MDN model accurately captures the standard
coding time required to make different kinds of code
change.

\begin{figure}[t!]
  \centering
  \includegraphics[width=0.99\linewidth]{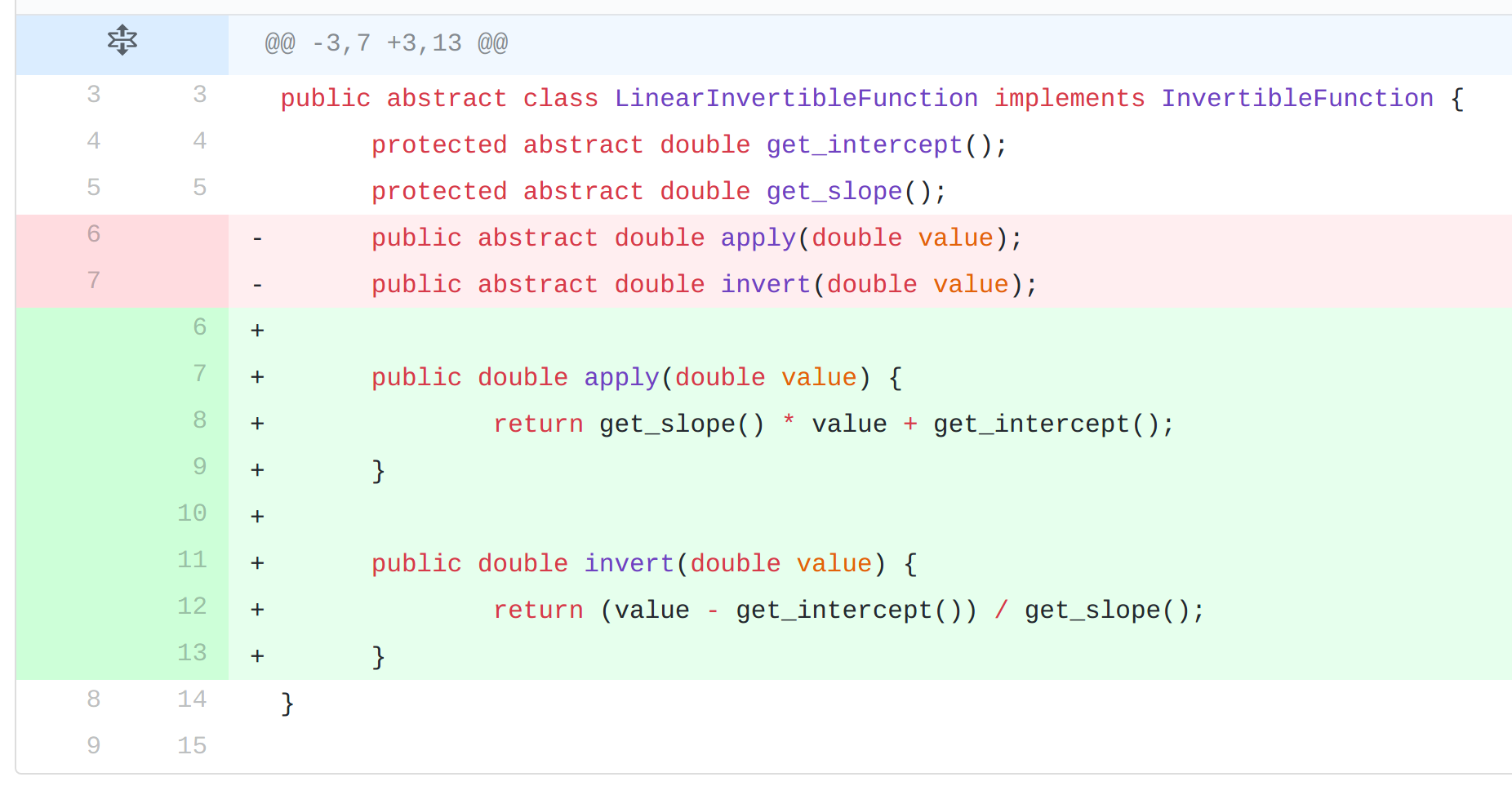}
  \caption{\small {\em Measuring the effort of a specific code change}. A developer deletes two and adds eight lines of code to one Java file. We construct a feature vector for this code change and query the MDN model. It predicts a mixture of 20 Gaussian distributions with $\mu=0.123$ (which means that the standard coder on average takes 7.38 minutes to make this type of change). Hence the effort represented by this code change is 0.123 Standard Coding Hours (SCH).}
  \label{fig:standardHours}
\end{figure}

In consequence, we can now measure the developer effort 
required to make specific code changes in terms of hours supplied by the standard coder, or Standard Coding Hours (SCH) (see Figure \ref{fig:standardHours}).

\subsection{Validation of the standard coder}

We can use the coding-time model (neural HMM) and the standard coder 
(deep MDN) to investigate the typical commits for open source projects. They use orthogonal inputs (commit timestamps and code deltas respectively) and measure different aspects of the `size' of a commit (how long it took to code and how long it should normally take, respectively). These different aspects are not perfectly identical since different projects vary in coding efficiency, due to factors such as domain complexity, project maturity and developer skill. 

\begin{figure}
  \centering
  \includegraphics[width=0.98\linewidth]{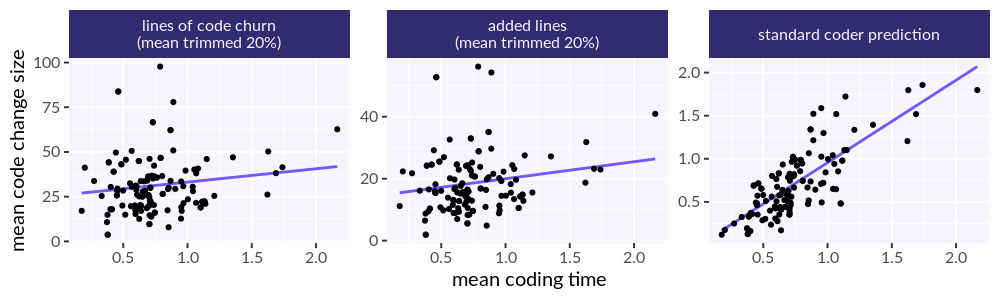}
  \caption{\small {\em Code change as progress per unit time}: For each project in our data set, consider the average time spent coding for one commit in this project versus the average `size' of the produced code change. We consider only commits from the projects' main contributors, and only projects with a sufficient number of such commits (500).}
  \label{fig:coding_input_vs_coding_output}
\end{figure}

Yet, in general, a project that spends more time on individual commits will have noticeably `larger' commits. If we measure the `size' of a commit
by new lines of code or 
churn (lines of code added plus lines of code deleted)
we observe only a very weak connection (Pearson coefficients $r = .25$ and $r = .21$ respectively for the 108 open source Java projects with more than 500 test commits from main developers; Spearman coefficients are lower).
The connection is however much stronger when using the standard coder to measure the commit `size' (Pearson coefficient of $r=0.80$, linear interpolation line of slope 0.98).\footnote{Since the standard coder provides a probability distribution, aggregation has been performed using unsupervised calibration \cite{unsupervisedCalibration}. However, the naive aggregation using means does not lead to a significantly different Pearson coefficient in this case either ($r=0.81$).} Figure \ref{fig:coding_input_vs_coding_output}
shows the superiority of the standard coder
over naively counting lines of code change.

This finding allows us to upper bound the variation in coding time due to differences in efficiency. The unexplained variance in Figure \ref{fig:coding_input_vs_coding_output} consists of differences in efficiency plus imperfections of our model. We can conclude that, in popular open source projects, differences in programming efficiency
account for at most $1-0.80^2 = 36\%$ of the variation in coding time. Note that this only concerns coding time, not time spent planning etc. 

\section{What kind of code changes require more or less effort?}

\subsection{Lines of code as a measure of quantity}

Our starting point was that metrics such as lines of code change are an overly simplistic measure of effort. We can now confirm that, while they are indeed statistically significantly correlated with our timestamp-based estimate of the time spent coding ($p < 0.1\%$), that correlation is neither strong nor linear. The Spearman correlation coefficients are listed in the left hand side of Table \ref{tab:simplistic_measure_correlations}. They demonstrate that the standard coder is vastly superior to surface criteria like lines of code.

The right hand side of Table \ref{tab:simplistic_measure_correlations} shows that the surface criteria listed, which are 
input features to the standard coder, do indeed influence the coding time, but explain only a part of its variation.

\begin{table}[]
    \centering
    \begin{tabular}{c|c|c}
        correlation with: & expected coding time & \begin{tabular}{c}standard coder\\ prediction\end{tabular}\\
        \hline
        files touched & 0.136 & 0.325\\
        spaces & 0.146 & 0.409\\
        tokens & 0.157 & 0.428\\
        lines added + deleted & 0.175 & 0.457\\
        lines added & 0.192 & 0.496\\
        standard coder prediction & 0.390 & 1.000\\
    \end{tabular}
    \caption{\small Spearman correlations between different predictors and coding time on the test set. All are statistically significant at a significance level of 0.1\%.}
    \label{tab:simplistic_measure_correlations}
\end{table}

Lines of code churn consists of added lines and deleted lines, where we treat modifications as deleting one line and adding another. Both the number of added lines and the number of deleted lines are correlated with coding time, but they are also correlated with each other. Once that relationship has been removed, it appears that the number of deleted lines alone does not positively contribute to the coding time: when maximising the Spearman coefficient between coding time and $\#(\mathrm{added\ lines}) + \beta\cdot\#(\mathrm{deleted\ lines})$ with a grid search of $\beta\in[-1, 1]$ with $\mathrm{step size} = 0.005$, this maximum is achieved for $\beta = -0.005$. This is reflected in the standard coder, whose highest correlation with the weighted sum of the number of added and deleted lines is likewise achieved at $\beta = -0.005$. 
Essentially, deleting old code appears to be of negligible effort compared to writing new code.

\subsection{Spread out or concentrated changes}

\begin{figure}
  \centering
  \includegraphics[width=0.9\linewidth]{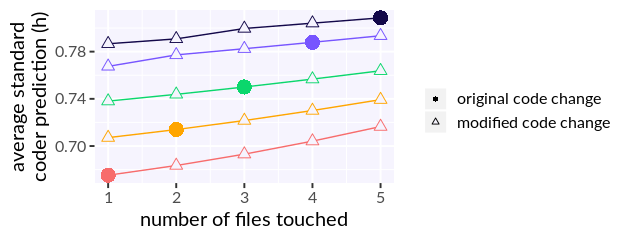}
  \caption{\small {\em More files touched more effort?} For each code change (with sufficient lines-of-code churn) we ask the standard coder how its prediction would change if the changes were distributed across a different number of files.}
  \label{fig:lines_of_code_effort}
\end{figure}

Modifying three lines in one file is different to modifying one line in three files each. For each code change in the corpus, we ask the standard coder what would happen if that change had been distributed across a different number of files.\footnote{We exclude code changes where the number of lines-of-code churn makes it clear that they could not be spread around the maximum number of files considered.} As shown in Figure \ref{fig:lines_of_code_effort}, the standard coder thinks code changes are easiest when concentrated in few files. However, this effect is dwarfed by code changes concentrated in few files being, on average, smaller in any case. On average, distributing otherwise identical code change over one more file increases the standard coding time by 32 seconds ($p < 0.1\%$ according to a paired Wilcoxon test). 

While the average effect looks roughly linear in Figure \ref{fig:lines_of_code_effort}, the individual impact of changing the number of files touched depends strongly on the individual code change. The interquartile ranges of the increase in
the standard coder's prediction are -5 to +56 seconds. The increase appears strongest for large commits concentrated in few files (for example, the average delta between one and two files changed is 47 seconds, and if in addition the starting effort is above the corpus median, the average delta is even 56 seconds). It appears that the more complicated a code change already is, the more effort it is to keep track of interactions between different files.

\subsection{Easy and hard concepts}

Querying the standard coder yields insight into which concepts are relatively easy or hard for humans to code. For example, all other things being equal, the same code change applied to a \code{public} method might take more time than when applied to a \code{private} method (since the programmer must try to consider all possible connections to other classes). Provided this is the case, and if the standard coder has learned this fact, then
we expect a code change with the keyword \code{public} to take more standard coding time 
compared to the same code change with the keyword \code{private}. We can ask similar questions
for other kinds of counter-factual keyword swaps
(e.g., we might expect that \code{implement}ing an interface is easier than \code{extend}ing a class).

However, we need to control for correlations between
keywords and the code context (e.g., 
\code{private} functions tend to be different from \code{public} functions) and the fact that 
mixtures of $n_A$ instances of \code{A} and $n_B$ instances of \code{B} are often considered more effortful than either $n_A + n_B$ instances of some keyword \code{A} or of \code{B}.

Therefore, to compare tokens \code{A} and \code{B}, we calculate, for each code change in our corpus that touches instances of \code{A} but not \code{B}, the change in the standard coder's prediction when we switch the {\em features} for \code{A} and \code{B}. We call the average of these deltas the \textit{cost} of B versus A and assess whether it is systematically higher than the cost of \code{A} versus \code{B} using a bootstrapped significance test. 

Table \ref{tab:cost_of_A_vs_B} demonstrates the results. In particular we find that:
\begin{enumerate}
    \item \code{public} methods and classes appear more difficult than \code{private} methods (with \code{protected} in between). This is an expected result, as mentioned above.
    \item Comparisons involving inequality appear more difficult than comparisons extending equality. This aligns with results from cognitive psychology that
    show that negative comparisons are inherently harder (\cite{negative_comparisons_1}, \cite{negative_comparisons_2}).
    \item Writing and extending classes is easier than writing and implementing interfaces (with the same code). This did not align with our previous expectations.
\end{enumerate}

\begin{table}[]
    \centering
    \begin{tabular}{ccc|c|c}
         \multicolumn{3}{c|}{tokens swapped} & cost &  \begin{tabular}{c}statistical\\ significance\end{tabular}\\
        \hline
        \hline
        private& to &public & 83s & \multirow{ 2}{*}{p < 1e-03}\\
        public& to &private & -54s & \\
        \hline
        private& to &protected & 22s & \multirow{ 2}{*}{p < 1e-03}\\
        protected& to &private & -33s & \\
        \hline
        protected& to &public & 24s & \multirow{ 2}{*}{p < 1e-03}\\
        public& to &protected & -50s & \\
        \hline
        <=& to &< & 34s & \multirow{ 2}{*}{p < 1e-03}\\
        <& to &<= & -125s & \\
        \hline
        >=& to &> & 80s & \multirow{ 2}{*}{p < 1e-03}\\
        >& to &>= & -216s & \\
        \hline
        ==& to &!= & 20s & \multirow{ 2}{*}{p < 1e-03}\\
        !=& to &== & -10s & \\
        \hline
        interface& to &class & -38s & \multirow{ 2}{*}{p < 1e-03}\\
        class& to &interface & -8s & \\
        \hline
        implements& to &extends & -4s & \multirow{ 2}{*}{p < 1e-03}\\
        extends& to &implements & 8s & \\
    \end{tabular}
    \caption{\small The average effect, according to the standard coder, of swapping all tokens of a certain kind that is present in a code diff for another kind not originally present in the code diff. Significance was tested by bootstrapping with 100000 samples.}
    \label{tab:cost_of_A_vs_B}
\end{table}

\section{Threats to validity}

We used the portfolio of open source software development projects on LGTM.com, representing highly starred projects from different open source development platforms. We rely on this dataset being representative for software development in general with respect to both the relationship between code change and time spent coding, and between time spent coding and time elapsed between commits. We assume that the coding time to produce a code change is typically supplied during commit intervals. Widespread
deviation from this assumption -- for example developers
who simultaneously work on many different branches
and interleave their time between them without
committing to version control when switching  --
would considerably bias our data set.
For simplicity, we assume consecutive commits separated by more than 2 minutes consist of distinct and consecutive pieces of work, while commits separated by less than 2 minutes are squashed before processing. High correlations between specific kinds of code change and
sub 2 minute commits would considerably bias
our data set. Finally, we rely on the inductive
bias of the coding time model (in particular the Markov assumption) being roughly sensible.

\section{Conclusion}

The metre is a standard measure of length. The standard coder is an analogous `measuring rod' that measures the effort required to make code changes in Standard Coding Hours (units of coding hours supplied by the standard coder). We therefore reduce heterogeneous, structured code changes to a scalar measure of effort. Importantly, the standard coder represents the actual living practice of software developers as manifested in large volumes of version control data. 
The standard coder therefore yields, in principle, a maximally unbiased measure of effort based on empirical regularities. A specific code change is high (or low) effort because the community of software developers typically requires more (or less) effort to produce this kind of code change.


The standard coder is a significantly better predictor of coding effort compared to simply counting lines of code and, in addition, yields insights about the difficulty of different coding tasks (e.g., changes across more files require more effort, inequalities are harder than equalities etc.)
However, the methodological approach of constructing a standard coder is more important than our particular model. We can expect to gain deeper insights by refining our model type, extending to other programming languages, and scaling-up to larger datasets.



\end{document}